\documentclass[10pt,letterpaper]{article}
\usepackage[top=0.85in,left=2.75in,footskip=0.75in]{geometry}

\usepackage{amsmath,amssymb}

\usepackage{subcaption}

\usepackage{changepage}

\usepackage[utf8x]{inputenc}

\usepackage{textcomp,marvosym}

\usepackage{cite}

\usepackage{nameref,hyperref}

\usepackage[right]{lineno}

\usepackage{microtype}
\DisableLigatures[f]{encoding = *, family = * }

\usepackage[table]{xcolor}

\usepackage{array}

\newcolumntype{+}{!{\vrule width 2pt}}

\newlength\savedwidth

\raggedright
\usepackage{ragged2e}

\usepackage[aboveskip=1pt,labelfont=bf,labelsep=period,justification=raggedright,singlelinecheck=off]{caption}

\makeatletter
\renewcommand{\@biblabel}[1]{\quad#1.}
\makeatother

\usepackage{lastpage,fancyhdr,graphicx}
\usepackage{epstopdf}
\fancyheadoffset[L]{2.25in}
\fancyfootoffset[L]{2.25in}
\begin{document}
\vspace*{0.2in}

\begin{flushleft}
{\Large
\textbf\newline{Measurement and Control of Main Spatio-Temporal Couplings in a CPA Laser Chain} 
}
\newline

Adeline Kabacinski\textsuperscript{1,*},
Kosta Oubrerie\textsuperscript{1},
Jean-Philippe Goddet\textsuperscript{1},
Julien Gautier\textsuperscript{1},
Fabien Tissandier\textsuperscript{1},
Olena Kononenko\textsuperscript{1},
Amar Tafzi\textsuperscript{1},
Adrien Leblanc\textsuperscript{1},
Stéphane Sebban\textsuperscript{1},
Cédric Thaury\textsuperscript{1}
\\
\bigskip
\textbf{1} Laboratoire d’Optique Appliquée, ENSTA-Paristech, CNRS, Ecole Polytechnique, Institut Polytechnique de Paris, 828 Bv des Maréchaux,91762 Palaiseau, France
\\
\bigskip

* adeline.kabacinski@ensta-paris.fr

\end{flushleft}

\section*{Abstract}

\justifying
We report a straightforward method to control main spatio-temporal couplings in a CPA laser chain system using a specially designed chromatic doublet in a divergent beam configuration. The centering of the doublet allows for the control of the spatial chirp of the CPA laser chain, while its longitudinal position in the divergent beam enables the control of the amount of longitudinal chromatism in a wide dynamic range. The performance of this technique is evaluated by measuring main spatio-temporal couplings with a simple method, based on an ultrafast pulse shaper, which allows for a selection of narrow windows of the spectrum. 

\section{Introduction}

Chirped pulse amplification (CPA) laser chains are being more and more used as they provide applications in a wide range of domains, thanks to their ultrashort and intense pulses\cite{Nolte:16}. The beam properties in the spatial domain can be controlled thanks to deformable mirrors or spatial light modulators \cite{Weiner:00}, while acousto-optic programmable dispersive filters allows for a temporal and spectral shaping of ultrashort beam\cite{Verluise:00}. However, the control of spatio-temporal couplings (STCs), described in \cite{Akturk:10}, still represents a challenge in the shaping of ultrashort pulses. 

Indeed, high energy CPA laser systems require large beam expanders, which can introduce couplings, better known as pulse front curvature in the near field and longitudinal chromatism in the far field, deteriorating the spatio-temporal focusing performance\cite{Bor:89}. A precise and dynamic adjustment of the STCs would also pave the way for new advances. Indeed, numerical calculations show that the combination between longitudinal chromatism and temporal chirp allows for the control of the velocity of ultrashort pulses under vacuum\cite{Sainte-Marie:17}, \cite{Froula:18}, therefore offering perspectives for laser-driven acceleration of ions\cite{Macchi:13}, or to avoid velocity mismatch effects in experiments involving the propagation of multiple pulses of different frequencies in a dispersive medium\cite{Depresseux:15}. Adjustment of longitudinal chromatism, along with spatial-phase shaping, also opens up prospects for phase-locked laser wakefield acceleration of relativistic electrons\cite{Caizergues:20}, \cite{Froula:20}.
Longitudinal chromatism due to lenses can be entirely avoided by replacing them by reflective optics, like spherical or parabolic mirrors. Yet, the former often introduce spherical aberration and even astigmatism, when the latter are expensive and more difficult to implement. Several compensation schemes have been proposed using reflective, refractive, or diffractive optical components\cite{Madjidi-Zolbanine:79}, \cite{Piestun:01}, \cite{Bahk:14}, \cite{Neauport:07}. All these methods have their own advantages and drawbacks. However, they have in common the inability to adjust continuously the longitudinal chromatism.
A dynamic chromatic aberration pre-compensation scheme for ultrashort laser chains has been suggested\cite{Cui:19}. It consists in a system, requiring vacuum to avoid air breakdown, composed of two lenses and a spherical mirror, each being on a translation stage. Although being efficient and having a good dynamic, this system remains expensive, complex and bulky. 

Another STC is inherent to ultrashort laser chains: pulse compressors require the introduction of massive amounts of angular dispersion, i.e. angular variation of the average wavelength in the beam\cite{Gu:04}, which is equivalent to pulse front tilt in the Fourier space. Focusing a beam containing angular dispersion/pulse front tilt generates spatial chirp, i.e. variation of the focus transverse position with respect to the wavelength. Although supposed to be completely removed with a perfect alignment, small misalignments as well as optics imperfections, are likely to cause non-desired angular dispersion, once again affecting the spatio-temporal focusing performance of the beam. On the other hand, some applications, like femtosecond pulse shaping\cite{Weiner:88} or control of the emission angle of laser-plasma accelerated electrons\cite{Popp:10}, require to separate spatially the different wavelengths deliberately and in a controlled manner. This can easily be done in the dispersion direction of the compressor gratings by changing their parallelism. More generally, some solutions have been suggested, like placing a lens one focal length away from a grating\cite{Weiner:88}. But this proposition is only suitable to introduce a precise amount of spatial chirp, depending on the properties of the chosen grating. 

In this letter, we report a straightforward method to control STCs using a chromatic doublet, inserted in a divergent beam configuration. This technique is of high interest since it allows to adjust both spatial chirp thanks to the centering of the doublet, and longitudinal chromatism in a wide dynamic range. The doublet design can easily be adapted to all types of applications, from full correction of the STCs to a precise control of their amount. 

\section{Principle and design of the doublet}
\begin{figure}[htbp]
\centering
\includegraphics[width=12cm]{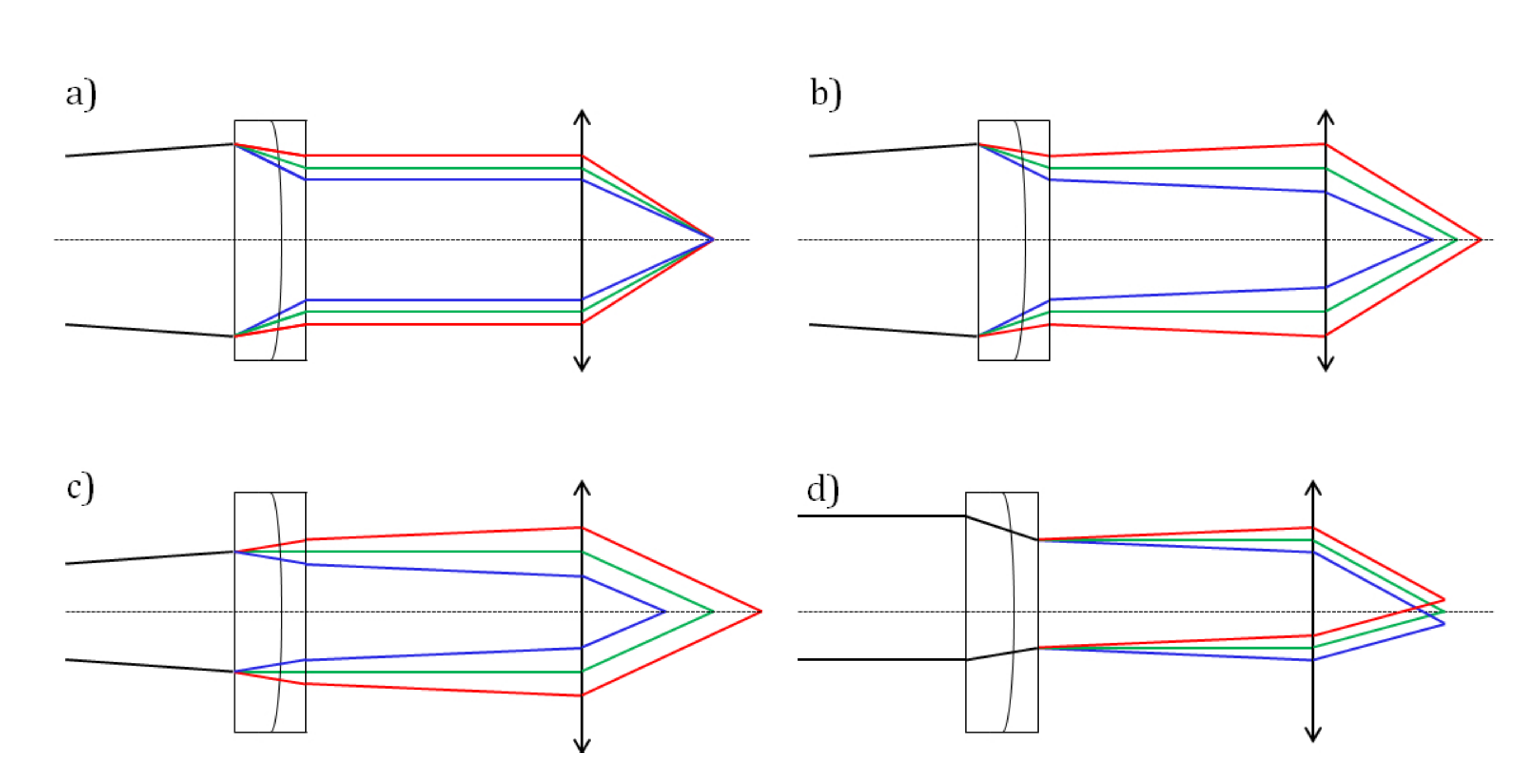}
\captionsetup{justification=justified}
\caption{Principle of the experiment. Rays plots for : a) a doublet designed to compensate exactly the longitudinal chromatism of the beam. b) and c) a doublet designed to introduce longitudinal chromatism for the same beam divergence but for a different longitudinal position of the doublet in this beam. d) a chromatic doublet off-center with respect to the beam. In d), the input beam is collimated to ease comprehension. A paraxial lens is added in each case to visualize the impact of the doublet.}
\label{fig:Principle}
\end{figure}

The control of the STCs (longitudinal chromatism and spatial chirp) is ensured by the properties of the doublet, and its longitudinal and transverse position in the beam. The principle on which this technique is based is as follows, and illustrated on figure~\ref{fig:Principle}. 

For decades, doublets have been used to reduce chromatic aberrations. They consist of two lenses, one convergent and one divergent, made of two different glasses which disperse wavelengths differently. Their properties are chosen to correct exactly the longitudinal chromatism. Therefore, there is no spreading of the focusing point along the beam propagation direction. Such an achromat is displayed on figure~\ref{fig:Principle} a) for a divergent beam. The represented doublet has an infinite focal length. 
On the contrary, doublets can be used to introduce a certain amount of longitudinal chromatism, characterized by the spectral focal shift (SFS), i.e. the distance between the respective focus of the considered wavelengths. The amount of longitudinal chromatism cannot be adjusted for a given doublet used in a collinear beam. However, this is not the case in a divergent beam as it can be seen on figure~\ref{fig:Principle} b) and c). They represent ray plots for an input beam with the same divergence, but for a different longitudinal position of the doublet in this beam. The focal planes of the different wavelengths are much closer to each other in b) than in c). Thus, using a doublet in a divergent beam allows for the adjustment of longitudinal chromatism. 
As shown on figure~\ref{fig:Principle} d), a decentering of the doublet induces angular dispersion and therefore a variation of the focus transverse position with respect to the wavelength. The more off-center the beam, the more the wavelengths are separated from each other in the transverse plane. Therefore, varying on the transverse (vertical and lateral) positions of the doublet, the amount of spatial chirp can be adjusted within the plane perpendicular to the propagation axis, and not only along one axis.

The doublet used for this experiment was specially designed for our laser chain using the software Zemax. The doublet is composed of a convergent lens in BK7 and of a divergent lens in SF5. The doublet is vacuum-spaced. The B integral of the doublet is estimated to be 3 mrad. It has an infinite focal length at 800 nm and is suited to a beam with a divergence of $\theta$ = 1.75°. More details concerning its design are given in the supplementary materials. It is conceived to introduce a variation of a few fs/cm$^{2}$ when translating the doublet longitudinally over a range of 40 cm. The doublet is optimized to minimize other aberrations. More details about the residual aberrations are provided in the supplementary materials. Moreover, moving the doublet by 40 cm along the optical axis only introduce a defocus of a few hundred of microns. Thus, the adjustment of the longitudinal chromatism can be performed without modifying the position of the focal plane.

The experiment was carried out using a 5-stages amplification Ti:Sa laser chain delivering 30fs-duration pulses with an energy up to 5J at a repetition rate of 1Hz. The spectrum of the laser is centered around 800 nm, with a spectral width of 150 nm at 1/$e^{2}$. The doublet is inserted in a divergent beam configuration, before the last lens of the last afocal system of the chain. The doublet is placed on a 40 cm translation mount in the longitudinal direction. Its axial position will therefore be expressed by the distance d between itself and the last lens of the last afocal system. The doublet is also motorized in vertical and lateral directions, to ease its centering and thus the adjustment of the spatial chirp. The measurement of longitudinal chromatism has been performed with a 60 mm beam diameter at the output of the last afocal system, while the measurement of spatial chirp has been performed with a 30 mm beam diameter. 

\section{Measurement and control of longitudinal chromatism}

\begin{figure}[htbp]
\centering
\includegraphics[width=\linewidth]{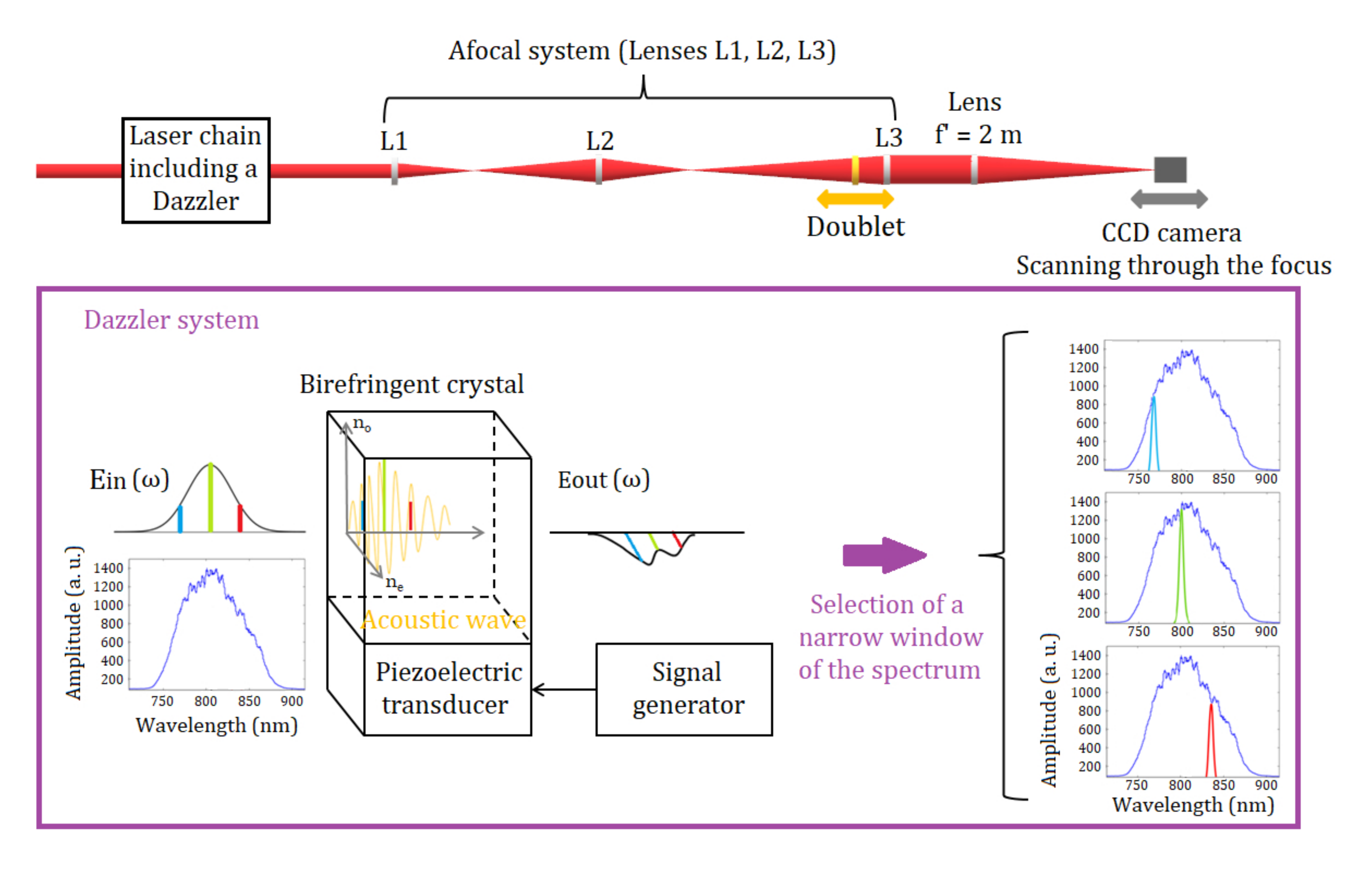}
\captionsetup{justification=justified}
\caption{Setup of the experiment. An ultrafast pulse shaper, installed upstream the amplification stages, selects a narrow window of the initial spectrum. The longitudinal focal plane, for a given spectral component, is located precisely with a CCD camera scanning through the focus of a lens of 2 m focal length. The doublet is placed on a 40 cm translation mount in the longitudinal direction and is also motorized in vertical and lateral directions. The measurements are performed for different distances between the doublet and the last lens of the afocal system. The whole laser chain is not represented, except its last afocal system.}
\label{fig:Setup}
\end{figure}

\begin{figure*}[ht]
\centering
\includegraphics[width=\linewidth]{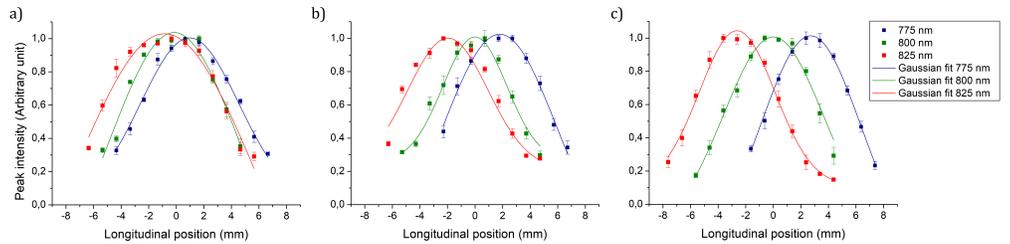}
\captionsetup{justification=justified}
\caption{Peak intensity of the beam with respect to the longitudinal position of the CCD camera for 775 nm in blue, 800 nm in green, 825 nm in red, a) for d = 12 cm, b) for d = 30 cm, c) for d = 50 cm.}
\label{fig:Scan}
\end{figure*}

The performance of our dynamic correction using the specially designed doublet was evaluated thanks to the following method, performed at reduced power to prevent damaging the amplifiers. An acousto-optic programmable dispersive filter (FastLite Dazzler ultrafast pulse shaper) installed upstream the amplification stages, is used to select a narrow window of the spectrum of width $\Delta\lambda$ = 10 nm. In practice, the incident beam propagates through a birefringent crystal, mounted on a piezoelectric transducer which creates a longitudinal traveling acoustic wave. The diffraction efficiency for each phase-matched optical spectral component is adjusted by modulating the amplitude of the traveling acoustic wave. By modulating its frequency, the longitudinal position where the diffraction occurs in the crystal for each spectral component can be tuned. Thus, controlling the spectral phase and the amplitude of the incident beam, certain components can be attenuated, resulting in the selection of a narrow window of the spectrum. The beam is then focused by a lens of 2 m focal length. With a CCD camera scanning through the Rayleigh zone of this lens, the longitudinal position of the focal plane for a given spectral component can be precisely located. For this measurement, three narrow windows, respectively centered on 775, 800 and 825 nm, were used. The setup of the experiment is shown on figure~\ref{fig:Setup}. The whole laser chain is not represented, except its last afocal system. The measurements are performed for different values of the distance d between the doublet and the last lens of the last afocal system.

The results of these measurements are displayed on figure~\ref{fig:Scan} for several values of d : a) 12 cm, b) 30 cm, c) 50 cm. It shows the peak intensity of the beam with respect to the longitudinal position of the CCD camera for 775 nm in blue, 800 nm in green and 825 nm in red. The measurement points are represented by squares, with error bars calculated as the standard error determined by repeatability over 5 shots. These points are fitted by Gaussians, represented in solid lines.

The performance of the doublet is evaluated through measurement of the shift of the focal plane for different wavelengths, called spectral focal shift (SFS). The obtained values are listed in the table~\ref{Table:Values}. The uncertainties are evaluated considering the standard error of the mean value of each fit. 

Therefore, our specially designed doublet allows for the control of the longitudinal chromatism in a wide dynamic range. In our configuration, a 40 cm range on the position of the doublet induces a linear variation of the SFS between 775 and 825 nm from 1.81 mm to 5.33 mm, which corresponds to a variation of pulse front curvature from $\alpha$ = 1.26 fs/cm$^{2}$ to 3.70 fs/cm$^{2}$. This range can be settled with a different design of the doublet, and thus be adjusted to all kind of applications, from a complete correction of the longitudinal chromatism for a chromatic laser chain to a precise control of the amount of longitudinal chromatism, wether in positive or in negative. 

\begin{table}
\begin{center}
\begin{tabular}{|c||c|c|c|}
  \hline
      & SFS 775-800 nm & SFS 775-825 nm  \\
  \hline
  d = 12 cm & (1.12 $\pm$ 0.18) mm & (1.81 $\pm$ 0.19) mm \\
  d = 30 cm & (1.80 $\pm$ 0.19) mm & (3.76 $\pm$ 0.25) mm \\
  d = 50 cm & (2.70 $\pm$ 0.11) mm & (5.33 $\pm$ 0.10) mm \\
  \hline
\end{tabular}
\end{center}
\captionsetup{justification=justified}
\caption{Measured values of the spectral focal shift between 775 and 800 nm, and of the spectral focal shift between 775 and 825 nm for different values of d.}
\label{Table:Values}
\end{table}

\section{Spatial chirp control}

\begin{figure}[htbp]
\centering
\includegraphics[width=7cm]{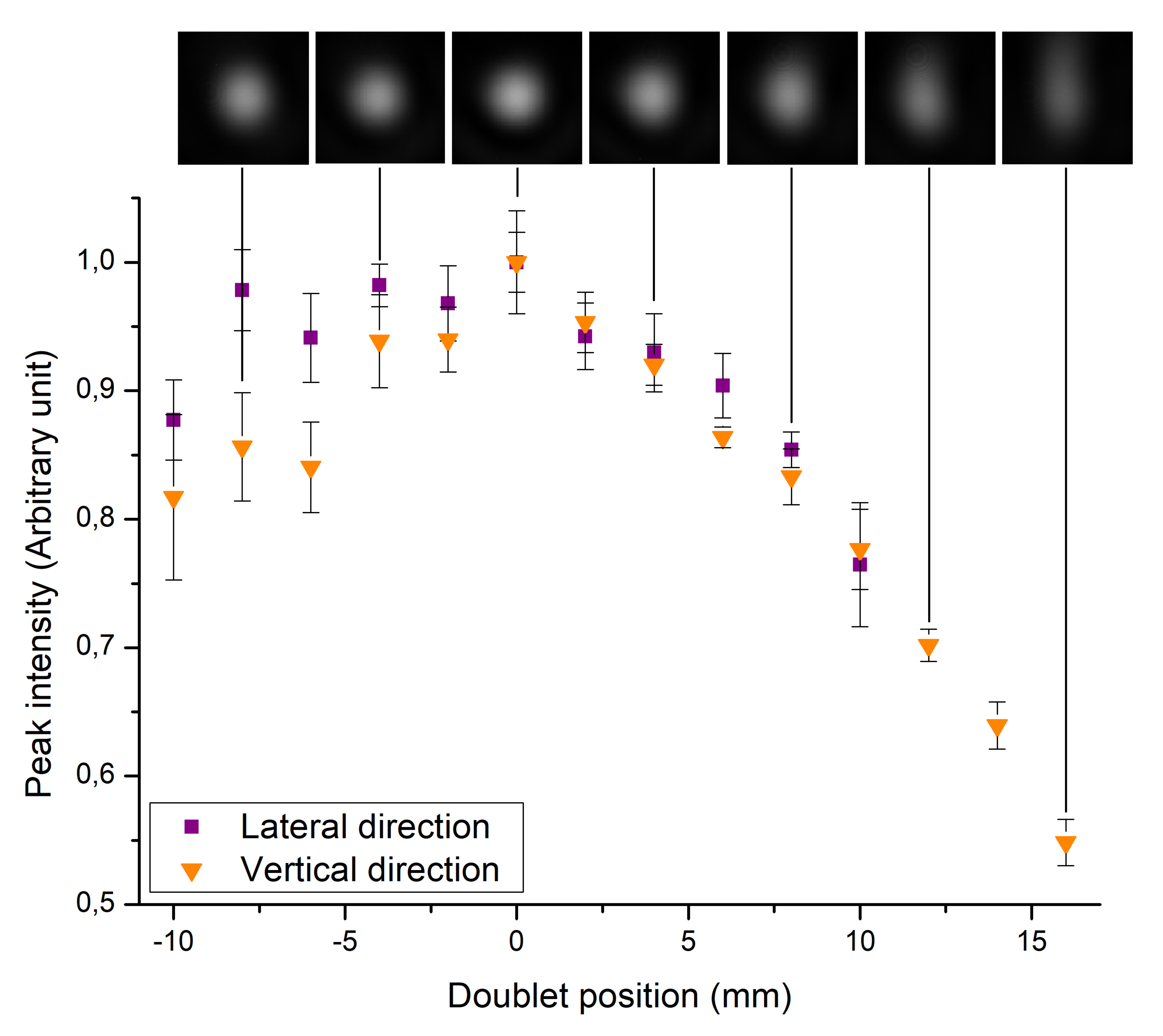}
\captionsetup{justification=justified}
\caption{Peak intensity at focus with respect to the transverse position of the doublet. Purple stands for the horizontal direction, orange for the vertical direction. At the top, spatial profiles of the beam for different vertical doublet positions.}
\label{fig:ChirpAllSpectrum}
\end{figure}

\begin{figure}[htbp]
\centering
\includegraphics[width=6cm]{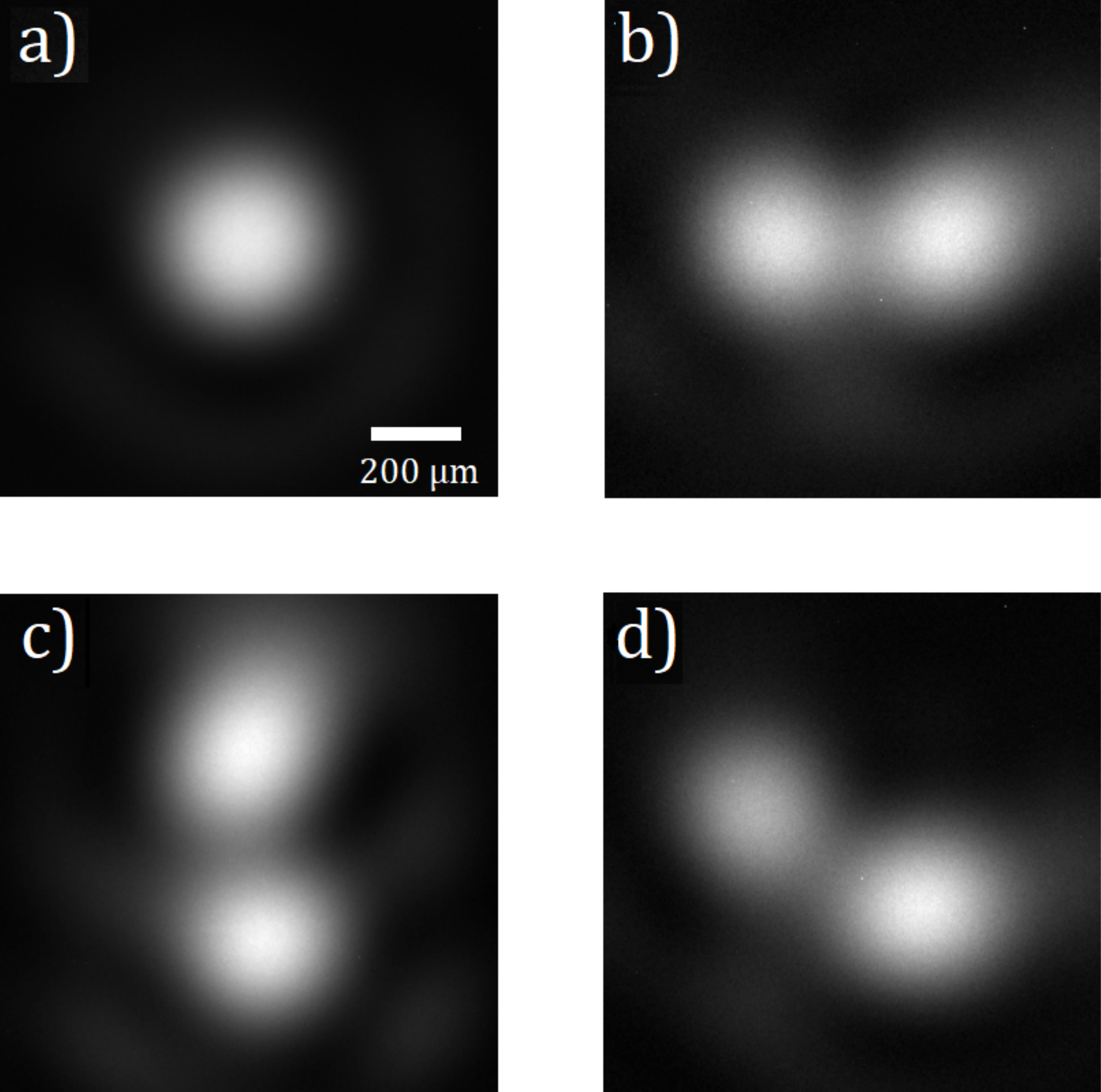}
\captionsetup{justification=justified}
\caption{Spatial profiles of two-colour laser beam pulse at the focus: a) for a perfect centering of the doublet, b) for a perfect vertical centering and a lateral decentering of 13 mm, c) for a perfect lateral centering and a vertical decentering of 15 mm, and d) for a vertical decentering of 7 mm and a lateral decentering of 13 mm. The beam radius at 1/e² on a) is of the order of 200 $\mu$m.}
\label{fig:Chirp2wavelengths}
\end{figure}

The use of this doublet also enables the correction of another STC. To visualize the impact of the centering of the doublet on the spatial chirp, we measure the peak intensity of the beam at focus with respect to the transverse position of the doublet. It is represented on figure~\ref{fig:ChirpAllSpectrum}, in purple for the lateral direction and in orange for the vertical direction. This figure enhances a maximum peak intensity for a perfect centering of the doublet, marked by position 0 in both directions. For each direction, there is a clear decrease of intensity when the doublet is not perfectly centered. Indeed, as it has been shown on figure~\ref{fig:Principle}, a decentering of the doublet introduces spatial chirp which results in a separation of the different frequency components at focus in the transverse directions. This can easily be visualized at the top of figure~\ref{fig:ChirpAllSpectrum}, showing spatial profiles of the beam at focus for different vertical doublet positions. For a perfect centering, the profile is very intense and circular, while it stretches in the direction of the decentering due to the separation of wavelengths.

Another way to display the spatial chirp consists in using again the ultrafast pulse shaper to adjust the spectrum so that it remains only two peaks, in this case at 775 and 825 nm respectively, each with a width $\Delta\lambda$ = 10 nm. A similar method is used in \cite{Borzsonyi:12} to measure spatial dispersion by filtering spectrally the beam to create well separated peaks in the spectrum. Figure~\ref{fig:Chirp2wavelengths} shows spatial profiles of the beam at the focus for different decentering of the doublet : a) was obtained for a perfect centering of the doublet, b) for a perfect vertical centering and a lateral decentering of 13 mm, c) for a perfect lateral centering and a vertical decentering of 15 mm, and d) a vertical decentering of 7 mm and a lateral decentering of 13 mm. The maximum decentering shown here, 15 mm in the vertical direction, causes an spatial chirp of 3.2 $\mu$rad/nm. For comparison, the spatial chirp required in \cite{Popp:10} was 1$\mu$rad/nm to deflect the electron beam by 4 mrad. Such a decentering of the doublet introduces small aberrations : simulations performed with Zemax showed that a decentering of 13 mm in lateral and 7 mm in vertical introduces 0.05 $\lambda$ of 45° astigmatism. As this value does not vary with the wavelength, it can easily be corrected with a deformable mirror. 

Varying the vertical and lateral position of the doublet, the position of one wavelength relatively to the other can easily be selected, and thus the amount of spatial chirp can be adjusted. This should be of high interest for pulse shaping experiments as the accuracy of pulse shaping depends on the degree of spatial chirp at the focal plane.

\section{Summary}
In summary, we have reported a straightforward method to evaluate the longitudinal chromatism of a CPA laser chain, using an ultrafast pulse shaper to locate precisely the focal plane of different spectral components. This technique is easy-to-implement since it only requires an acousto-optic dispersive filter, available in most laser facilities, and a focusing optic. This process allowed us to assess the performance of a chromatic doublet regarding the control of STCs. The longitudinal position of the doublet, inserted in a divergent beam configuration, enables the regulation of the amount of longitudinal chromatism in a wide dynamic range. This technique can easily be adapted to most laser chains by adjusting the design of the doublet. Furthermore, the centering of the doublet allows for the control of the degree of spatial chirp at focus. It opens up prospects for numerous applications requiring the adjustment of STCs.

\section*{Acknowledgments}
This work has received funding from the European Union’s Horizon 2020 Research and Innovation programme under Grant No. 730871 (project ARIES).

\newpage

\section*{SUPPLEMENTAL DOCUMENT}

\section*{Design of the doublet}

\subsection*{Characteristics}
\begin{figure}[htbp]
\centering
\fbox{\includegraphics[width=.6\linewidth]{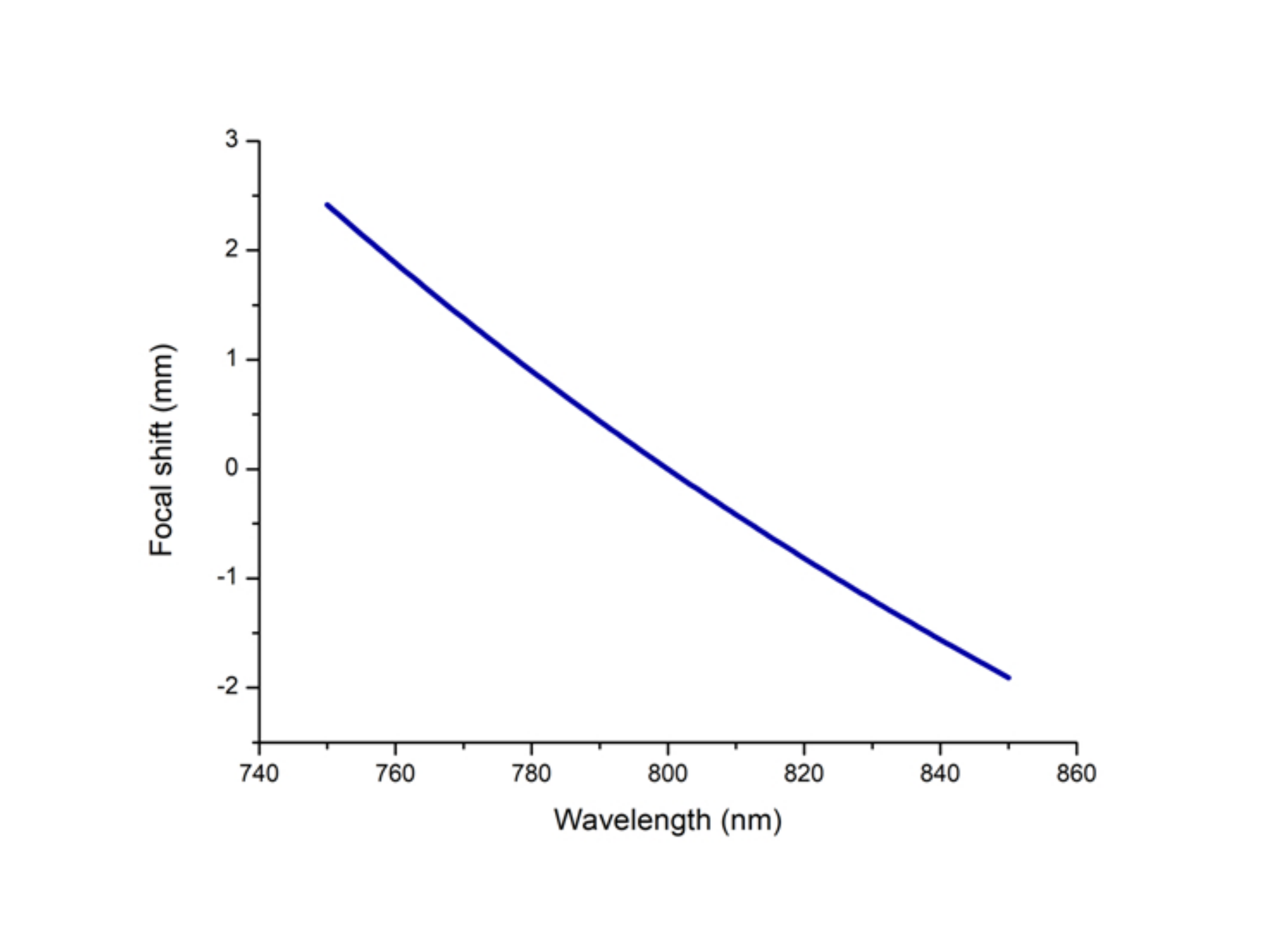}}
\caption{Focal shift of the doublet}
\label{fig:Doubletfocalshift}
\end{figure}

The doublet is composed of a convergent lens (lens 1) in BK7 and of a divergent lens (lens 2) in SF5, spaced by vacuum. It has an infinite focal length at 800 nm. The focal shift of the doublet is shown on figure \ref{fig:Doubletfocalshift}. The main doublet characteristics are given in the table \ref{tab:Characteristics}.

\begin{table}[htbp]
\begin{center}
\begin{tabular}{|c||c|}
  \hline
  Doublet diameter & 100 mm\\
  \hline
  Curvature radius, lens 1 & 736 CC mm and 210.6 CX mm\\
  \hline
  Curvature radius, lens 2 & 214.67 CC mm and 499.8 CX mm\\
  \hline
  Thickness, lens 1 & 4.68 mm\\
  \hline
  Thickness, lens 2 & 10.4 mm\\
  \hline
  Thickness, vacuum &  2 mm\\
  \hline
\end{tabular}
\end{center}
\captionsetup{justification=justified}
\caption{Main doublet characteristics}
\label{tab:Characteristics}
\end{table}

\subsection*{Non linear phase shift}
A generally accepted criterion for high-power laser systems is that the cumulative B integral must be kept somewhere below B < 3 to 5 radians to avoid serious nonlinear damage and distortion effects due to either self-phase modulation or self-focusing. The B-integral of our laser chain in the full power mode is equal to 1.7 radians. Considering a non linear index of 3.05 $10^{-20}$ m$^{2}$/W for the BK7 and of 12.7 $10^{-20}$ m$^{2}$/W for the SF5, the B integral of the doublet is estimated to be around 3 mrad.

\section*{Aberrations introduced by the doublet}

\subsection*{Perfectly aligned doublet}
The doublet is optimized to minimize aberrations. The strongest aberration with a perfectly centered doublet is first order spherical aberration, which varies from -2.$10^{-3}$ $\lambda$ to 5.$10^{-3}$ $\lambda$ when we translate the doublet by 40 cm longitudinally in a beam of divergence $\theta$=1.75°.

\subsection*{Off-centered doublet}
Simulations performed with Zemax showed that a decentering of 13 mm in lateral and 7 mm in vertical introduces 0.05 $\lambda$ of 45° astigmatism. As these aberrations do not vary with the wavelength, it can easily be corrected with a deformable mirror. 

\subsection*{Tilted doublet}
For an extreme tilt value of 2°, simulations performed with Zemax show 0.2$\lambda$ of astigmatism and 0.2$\lambda$ of coma. 

\section*{Effect of the doublet on second and third phase order}

\begin{figure}[htbp]
\centering
\fbox{\includegraphics[width=0.9\linewidth]{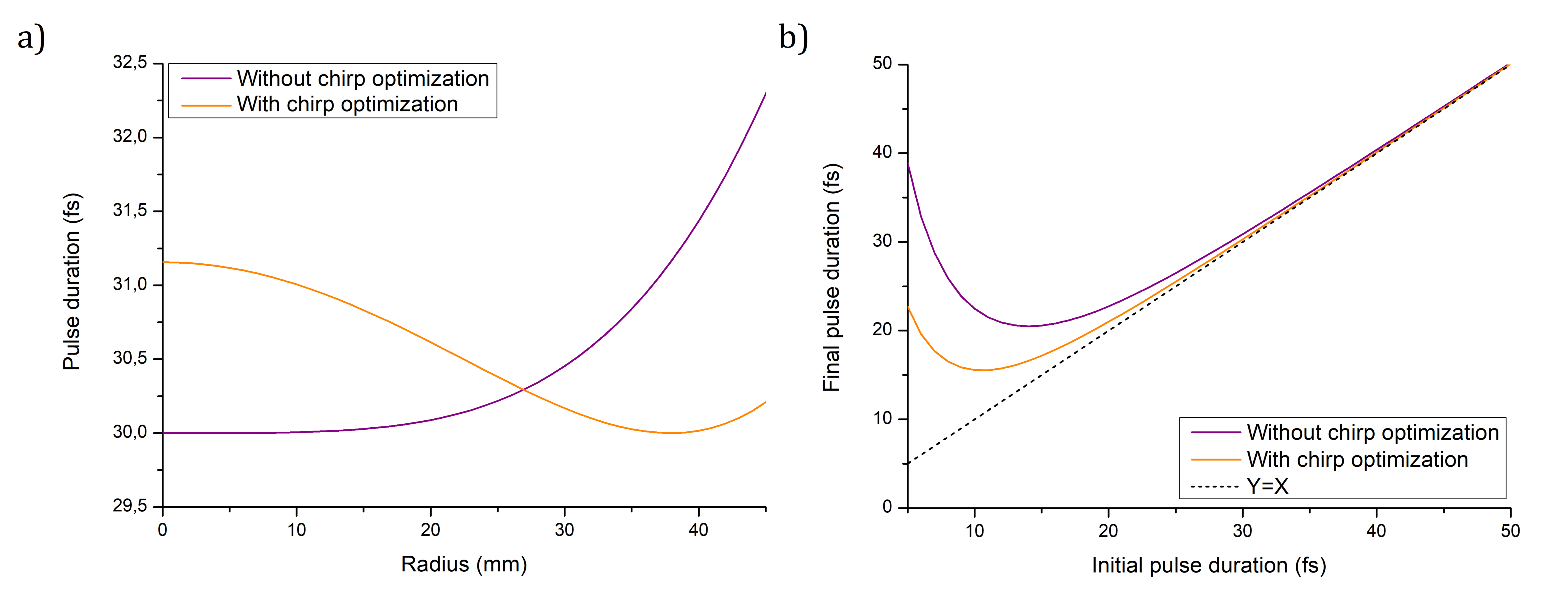}}
\caption{Effect of the doublet on second phase order : a) Pulse duration after the doublet with respect to the beam radius for an initial pulse duration of 30 fs. b) Pulse duration averaged on the whole top-hat beam after the doublet, with respect to the initial pulse duration. Purple curve does not consider chirp optimization, while orange curve does. The curve Y=X is displayed in dotted lines for the sake of readability.}
\label{fig:GDD}
\end{figure}

Considering a group velocity dispersion of 44.6 fs$^{2}$/mm for the BK7 and of 126.4 fs$^{2}$/mm for the SF5, and the variation of lenses thicknesses with respect to the beam radius, we calculated the total group delay dispersion introduced by the doublet and its impact on pulse duration. Figure\ref{fig:GDD} a) shows the pulse duration after the doublet with respect to the beam radius, for an initial pulse duration of 30 fs. Purple curve does not consider chirp optimization, while orange curve does. Taking into account a top-hat beam of radius 45 mm, we obtain a mean pulse duration of 30.9 fs. This mean duration can be reduced by introducing a constant chirp with the ultrafast pulse shaper or the compressor itself. We obtain 30.3 fs with chirp optimization. Figure\ref{fig:GDD} b) shows the averaged pulse duration on the whole beam after the doublet with respect to the initial pulse duration. The curve Y=X is displayed in dotted lines for the sake of readability. We deduce that the pulse duration is not significantly affected by the doublet for initial pulse duration larger than 15 fs. 

Considering a third phase order of 83 fs$^{3}$/mm for the SF5 and of 32 fs$^{3}$/mm for BK7 and the respective thicknesses of the lenses, we obtain a total of 721 fs$^{3}$ at the center and a difference of 86 fs$^{3}$ between the center and the edge of the doublet. The constant part of the third phase order can be easily corrected by the Dazzler, while the radial varying part can be considered as negligible. 

\end{document}